\documentclass[useAMS,usenatbib]{mn2e} 
\usepackage{aas_macros}
\usepackage{graphics}
\usepackage{epsfig}  
\usepackage{natbib} 
\usepackage{color}
\usepackage{float}
\usepackage{amsmath}
\usepackage{times}
\usepackage{upgreek}
\usepackage[varg]{txfonts}
\newcommand{\Tab}[1]{Table~\ref{#1}}
\newcommand{\Sec}[1]{Section~\ref{#1}}
\newcommand{\Eq}[1]{Eq.(\ref{#1})}
\newcommand{\Fig}[1]{Fig.\ref{#1}}

\newcommand{\hMpc}{{\ifmmode{h^{-1}{\rm Mpc}}\else{$h^{-1}$Mpc}\fi}}
\newcommand{\hkpc}{{\ifmmode{h^{-1}{\rm kpc}}\else{$h^{-1}$kpc}\fi}}
\newcommand{\hMsun}{{\ifmmode{h^{-1}{\rm {M_{\odot}}}}\else{$h^{-1}{\rm{M_{\odot}}}$}\fi}}
\newcommand{\ltsima}{$\; \buildrel < \over \sim \;$}
\newcommand{\gtsima}{$\; \buildrel > \over \sim \;$}
\newcommand{\lsim}{\lower.5ex\hbox{\ltsima}}
\newcommand{\gsim}{\lower.5ex\hbox{\gtsima}}

\def\lesssim{\mathrel{\hbox{\rlap{\hbox{\lower4pt\hbox{$\sim$}}}\hbox{$<$}}}}
\def\gtrsim{\mathrel{\hbox{\rlap{\hbox{\lower4pt\hbox{$\sim$}}}\hbox{$>$}}}}
\newcommand{\nil}[1]{#1}

\title[Backsplash Galaxies in the Local Group]
      {The luminosities of backsplash galaxies in constrained simulations of the Local Group}
\author[Knebe et al.] 
{Alexander Knebe$^1$, Noam I Libeskind$^2$, Steffen R. Knollmann$^1$, Luis A. Martinez-Vaquero$^1$, \newauthor Gustavo Yepes$^1$, Stefan Gottl\"ober$^2$, Yehuda Hoffman$^3$
  \\
  $^1$Grupo de Astrof\'\i sica, Departamento de Fisica Teorica, Modulo C-15, Universidad Aut\'onoma de Madrid, Cantoblanco E-28049, Spain\\
  $^2$Astrophysikalisches Institut Potsdam, An der Sternwarte 16, D-14482 Potsdam, Germany\\
 $^3$Racah Institute of Physics, The Hebrew University of Jerusalem,
 Jerusalem 91904, Israel
  }

\begin{document}

\date{First draft}

\pagerange{\pageref{firstpage}--\pageref{lastpage}} \pubyear{2008}

\maketitle

\label{firstpage}

\begin{abstract}
  We study the differences and similarities in the luminosities of
  bound, infalling and the so-called backsplash \citep{Gill05}
  galaxies of the Milky Way and M31 using a hydrodynamical simulation
  performed within the Constrained Local UniversE Simulation (CLUES)
  project. The simulation models the formation of the Local Group
  within a self-consistent cosmological framework. We find that even
  though backsplash galaxies passed through the virial radius of their
  host halo and hence may have lost a (significant) fraction of their
  mass, their stellar populations are hardly affected. This leaves us
  with comparable luminosity functions for infalling and backsplash
  galaxies and hence little hope to decipher their past (and
  different) formation and evolutionary histories by luminosity
  measurements alone. Nevertheless, due to the tidal stripping of dark
  matter we find that the mass-to-light ratios have changed when
  comparing the various populations against each other: they are
  highest for the infalling galaxies and lowest for the bound
  satellites with the backsplash galaxies in-between. 

\end{abstract}


\section{Introduction}
\label{sec:introduction}
Ever since \citet{Klypin99s} and \citet{Moore99} pointed out that
dark matter simulations of cosmic structure formation lead to an excess of
subhaloes as compared to the number of observed (luminous) satellite
galaxies \nil{visibly surrounding the Milky Way (MW) and M31,} the industry for
simulating and studying substructure in cosmological (dark matter)
haloes has boomed. The tension has been marginally loosened with the
discovery of a substantial number of new ultra-faint satellites
galaxies in the Local Group thanks to the SDSS data \citep{SDSS-DR5}:
within the past couple of years the number of known MW and
M31 satellites has nearly doubled. And taking into account the
detection limits and the sky coverage of the SDSS survey we will most
certainly stumble across even more \nil{galactic satellites} in the
near future when, for instance, upcoming data from GAIA and panSTARRS
have been analysed.

As noted by several groups before \citep{Moore04, Gill05},
there exists a prominent population of galaxies that are found outside
the virial region of their host at the present day, but whose orbits
took them inside the virial radius at earlier times. While their
studies were based upon cosmological simulations of galaxy clusters, the existence of this ``backsplash population'' has also been
\nil{found} for MW-type objects \citep{Warnick08, Ludlow09}. This raises the
question whether or not (and how) one can distinguish infalling and
backsplash galaxies from each other. \citet{Gill05} suggested to use
the line-of-sight velocity distribution: as shown in their Fig.8 the
distribution of line-of-sight velocities of subhaloes relative to the
host (and convolved with the 2dF velocity uncertainty of 100 km/sec)
is different for the infalling and the backsplash population. However,
there may be a simpler way that does not involve spectroscopy. Since
backsplash satellites \nil{had, at one point in their orbit, a closer
  approach to the central galaxy than infalling satellites}, the tidal
influence of the host \nil{must have been} stronger for the backsplash
population then for infalling satellites. Could this difference in tidal
forces effect the initial stellar population (if existent), and can it
be used to discriminate between the two populations?  It has been
shown by \citet{Gill05} that backsplash galaxies loose on average 40\%
of their initial mass when grazing their host. But what about the
stellar content? As baryons are able to cluster more strongly in the
centre of the potential well the stars are also more centrally
concentrated. Therefore, will the cold baryonic component be safe from
tidal stripping when the backsplash galaxy (briefly) flies through its
host? This question is the major motivation for this work. We address the issue of separating the three types of galaxies
(bound satellites, backsplash and infalling) by means of luminosity
(and possibly mass) measurements only.

\section{The Simulations}
\label{sec:simulations}
In this Section we describe the simulations used throughout this study
and the methodology employed to identify host haloes and their
substructure.

\subsection{Constrained Simulations of the Local Group}
\label{sec:localgroup}

We use the same set of simulations already presented in
\citet{Libeskind10} and \citet{Knebe10a} and refer the reader to these
papers for a more exhaustive discussion and presentation of these
constrained simulations of the Local Group that form part of the CLUES
project.\footnote{\texttt{http://www.clues-project.org}} However, we briefly summarize their main properties here for clarity.

We choose to run our simulations using standard $\Lambda$CDM initial
conditions, that assume a WMAP3 cosmology \citep{Spergel07}, i.e.
$\Omega_m = 0.24$, $\Omega_{b} = 0.042$, $\Omega_{\Lambda} = 0.76$. We
use a normalization of $\sigma_8 = 0.73$ and a $n=0.95$ slope of the
power spectrum. We used the PMTree-SPH MPI code \texttt{GADGET2}
\citep{Springel05} to simulate the evolution of a cosmological box
with side length of $L_{\rm box}=64 h^{-1} \rm Mpc$. Within this box
we identified (in a lower-resolution run utilizing $1024^3$ particles)
the position of a model local group that closely resembles the real
Local Group \citep[cf.][]{Libeskind10}. This Local Group has then been
re-sampled with 64 times higher mass resolution in a region of $2
h^{-1} \rm Mpc$ about its centre giving a nominal resolution
equivalent to $4096^3$ particles giving a mass resolution of $m_{\rm
  DM}=2.1\times 10^{5}$\hMsun\ for the dark matter and $m_{\rm
  gas}=4.42\times 10^4$\hMsun\ for the gas particles. For more details
we refer to the reader to \citet{Gottloeber10}.

For this particular study we focus on the gas dynamical SPH
simulation, in which we follow the feedback and star formation rules of
\cite{Springel03}: the interstellar medium (ISM) is modeled as a two
phase medium composed of hot ambient gas and cold gas clouds in
pressure equilibrium. The thermodynamic properties of the gas are
computed in the presence of a uniform but evolving ultra-violet cosmic
background generated from QSOs and AGNs and switched on at $z=6$
\citep{Haardt96}.  Cooling rates are calculated from a mixture of a
primordial plasma composition. No metal dependent cooling is assumed,
although the gas is metal enriched due to supernovae
explosions. Molecular cooling below $10^{4} {\rm K}$ is also ignored.
Cold gas cloud formation by thermal instability, star formation, the
evaporation of gas clouds, and the heating of ambient gas by supernova
driven winds are assumed to all occur simultaneously.

\subsection{The (Sub-)Halo Finding}
\label{sec:halofinding}
In order to identify halos and subhaloes in our simulation we have run
the MPI+OpenMP hybrid halo finder \texttt{AHF}\footnote{\texttt{AMIGA}
  halo finder, to be downloaded freely from
  \texttt{http://www.popia.ft.uam.es/AMIGA}} described in detail in
\cite{Knollmann09}. \texttt{AHF} is an improvement of the \texttt{MHF}
halo finder \citep{Gill04a}, which locates local over-densities in an
adaptively smoothed density field as prospective halo centres. The
local potential minima are computed for each of these density peaks
and the gravitationally bound particles are determined. Only peaks
with at least 20 bound particles are considered as haloes and retained
for further analysis (even though we place a tighter constraint on the
number of particles for the present analysis, cf. below). We like to
stress that our halo finding algorithm automatically identifies
haloes, sub-haloes, sub-subhaloes, etc. For more details on the mode
of operation and actual functionality we though refer the reader to
the code description paper by \citet{Knollmann09}.

Subhaloes are defined as haloes which lie within the virial region of
a more massive halo, the so-called host halo.  We build merger trees
by cross-correlating haloes in consecutive simulation outputs. For
this purpose, we use a tool that comes with the \texttt{AHF} package called \texttt{MergerTree}, that follows each halo (either host
or subhalo) identified at redshift $z=0$ backwards in time. The
direct progenitor at the previous redshift is the object that shares the
most particles with the present halo \textit{and} is closest to it in
mass. Again, for more elaborate details we point to the reader to
\citet{Libeskind10}.

\subsection{Lighting up Subhaloes}
\label{sec:lightingup}

The stellar population synthesis model STARDUST \citep[see][and
references therein for a detailed description]{Devriendt99} has been
used to derive luminosities from the stars formed in our
simulation. This model computes the spectral energy distribution from
the far-UV to the radio, for an instantaneous starburst
of a given mass, age and metalicity. The stellar contribution to the
total flux is calculated assuming a Kennicutt initial mass function \citep{Kennicutt98}.

\section{Results}
\label{sec:results}
The prime target of this study is to find possible differences in the
properties of backsplash, bound and infalling galaxies with respects
to luminosity. We explicitly \nil{use the term} ``galaxies'' as we focus
solely on subhaloes with a luminous stellar component; all other
objects will be neglected for this particular investigation.  We
further only consider satellites of the (simulated) MW and
Andromeda (M31) galaxy; the subhaloes of both these host haloes will
be stacked in the subsequent plots presented here. In addition to the
requirement for subhaloes to contain stars we also apply a lower mass
cut of $M>2\times 10^{7}$\hMsun\ roughly corresponding to 100
particles in total (note that particles have different masses as they
represent dark matter, gas and stars).

\subsection{The Existence of Backsplash Galaxies}
\label{sec:existence}
\begin{figure}
\noindent
\centerline{\hbox{\psfig{figure=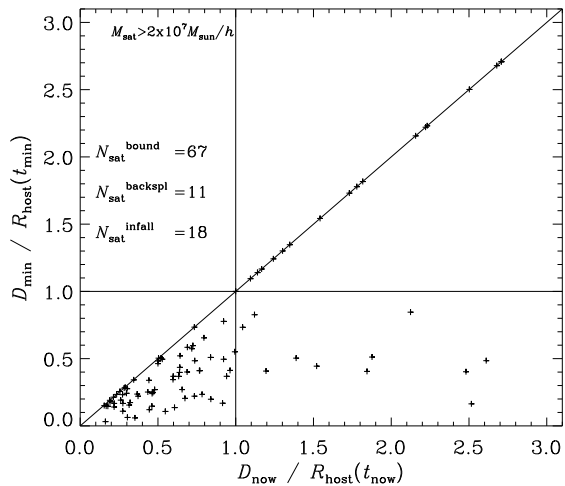,width=\hsize,angle=0}}}
\caption{Minimum distance $D_{\rm min}$ as a function of present-day
  distance $D_{\rm now}$ both normalized to the virial radius of the
  host at the respective time.}
\label{fig:backsplash}
\end{figure}

\begin{figure*}
\noindent
\centerline{\hbox{\psfig{figure=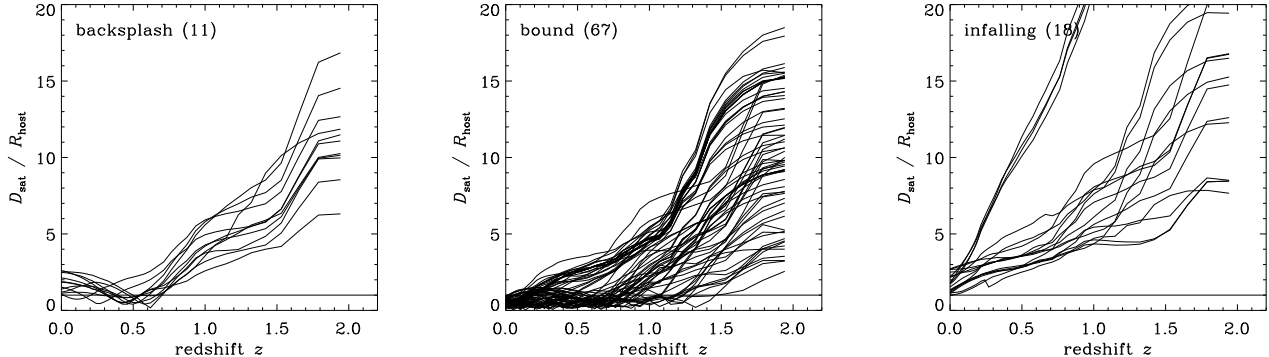,width=\hsize,angle=0}}}
\caption{The orbits of all considered subhaloes. The left panel shows the backsplash galaxies, the middle panel the bound and the right panel the infalling population.}
\label{fig:orbits}
\end{figure*}

Before examining the properties of backsplash galaxies we
\nil{wish to first} confirm their existence. To this extent we plot in
\Fig{fig:backsplash} the closest approach (normalized to the virial
radius of the satellite's host at the time of minimum distance)
vs. its present-day distance (normalized to its host's virial
radius). The number of objects in the respective population are given
in the legend. Note that we only plot those subhaloes that
contain a stellar component. This figure contains three
distinct parts defining the three different populations. First, those
subhaloes whose minimum distance equals its present-day distance are
the infalling population: they are continuously falling towards their
host. Second, galaxies that entered the virial radius of their host
and remained inside ever since. Even though the host radius is
increasing in size since the time a subhalo entered, we nevertheless
find that there are no subhaloes above the 1:1 line; we therefore
conclude that the increase in host radius as measured by $R_{\rm
  host}^{t_{\rm now}}/R_{\rm host}^{t_{\rm min}}$ is smaller than the
ratio $D_{\rm now}/D_{\rm min}$. This comes \nil{as} no surprise as we
\nil{do not} expect satellites to orbit on circular orbits ($D_{\rm
  now}=D_{\rm min}$); subhaloes \nil{may} have (highly) eccentric orbits taking
them close to the centre of their host \citep[cf. Figs.7 and 8
in][]{Gill04b}. Third and last, there are galaxies that once were
inside their host's virial radius but are presently found outside,
i.e. the backsplash population. 

\Fig{fig:backsplash} indicates that we might expect to find of order
40\% to be backsplash galaxies in the vicinity of the Milky Way and/or
M31 -- a percentage in agreement with previous studies of this class
of objects\citep[cf. ][]{Gill05, Warnick08}. The question now is
whether or not we will be able to distinguish these populations and
find the backsplash galaxies, respectively, by quantifying their
luminosities.

Before proceeding we \nil{would} like to add a cautionary remark
\nil{clarifying our terminology}: we \nil{call} subhaloes that are
inside their host's virial at $z=0$, ``bound''. Those \nil{subhaloes}
that \nil{were} once inside \nil{their hosts virial radius} but are
found at $z=0$ outside \nil{are termed} ``backsplash''. As we can see
from \Fig{fig:orbits} this classification strongly depends on the
redshift used to define the populations. We can see that a fair
fraction of today's bound population had been backsplashed in the
past, while probably all of today's backsplash galaxies will return
and re-enter their hosts \nil{at some future time}. Therefore, the
expression ``bound'' should not be taken literally \nil{(in terms of
  energy arguments)} but rather as a reference to satellites under a
prolonged influence of their host while ``backsplash'' refers to
satellites under brief influence of their host. Further, please
note that we require objects to exist both at redshift $z=1.5$ and
today to be part of our sample; there are also subhaloes that were
present at high redshift but got tidally disrupted and hence did not
survive.

In preparation of the investigation the luminosities in
\Sec{sec:luminosities} and baryon content in \Sec{sec:baryons}, we
\nil{wish to} find the time where all three populations were still
infalling \nil{so as} to verify the correctness of our tracking scheme
for subhaloes. To this extent we present in \Fig{fig:orbits} the
\nil{distance from the center of their hosts} of all backsplash (left
panel), bound (middle panel), and infalling (right panel) galaxies
\nil{found and identified at} redshift $z=0$. The orbits have been
normalized to the virial radius of the respective host (at redshift
$z$) of each galaxy and hence the solid line $D_{\rm sat}/R_{\rm
  host}=1$ marks the ``entry'' (and ``exit'') point \nil{of the
  satellite} into \nil{and out of} the host. While this figure
\nil{succinctly} demonstrate that backsplash galaxies clearly
\nil{entered and exited} their host (\nil{while} infalling
\nil{galaxies} have not yet \nil{crossed the virial radius}) it also
allows us to find that point in time in our simulation at which
\textit{all} populations were still infalling: \nil{this can be seen
  at a} redshift $z\geq 1.5$. We will return to this redshift later as
we expect the properties of galaxies to be drawn from the same
statistical distribution at that time: no galaxy has yet entered the
host (or left again) which may (or may not) have caused changes in the
internal properties and -- in particular -- the luminosities.

\subsection{The Luminosities}
\label{sec:luminosities}
\begin{figure}
\noindent
\centerline{\hbox{\psfig{figure=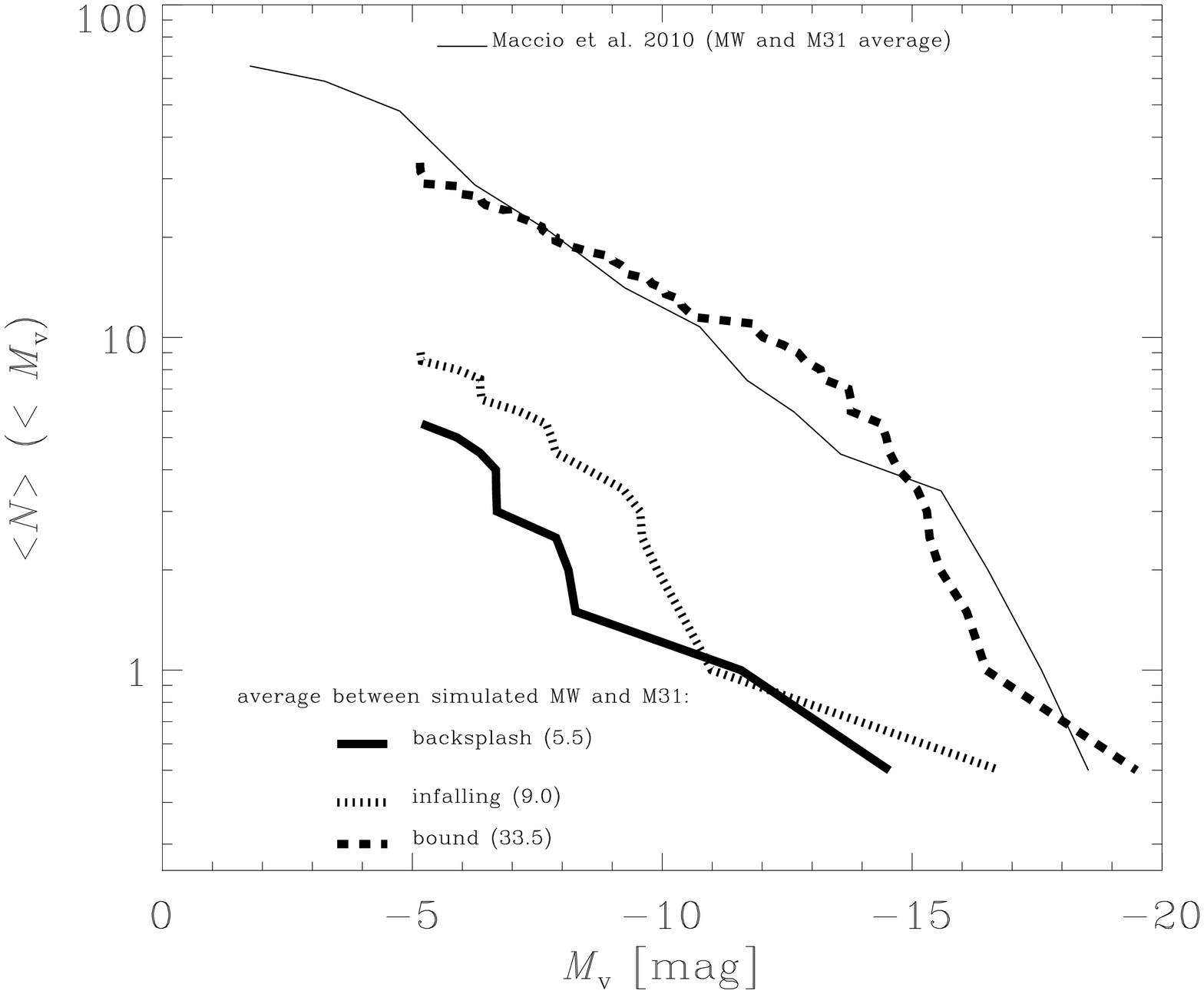,width=\hsize,angle=0}}}
\caption{The luminosity function of subhaloes in the Johnson
  V-Band. The ``Maccio'' observational data (thin solid line) is a
  combination of the volume corrected MW luminosity function
  \citet{Koposov08} augmented with information from \citet{Mateo98}
  and \citet{Maccio10} under the assumption of an NFW-like radial
  distributions of satellites. Note that the comparison to the
  observational data is \textit{not} the prime target of this study
  and only serves as a reference, respectively.}
\label{fig:luminosity}
\end{figure}
 
\nil{In this section we look at the luminosity of the stellar
  components of galaxies identified as bound, backsplash and
  infalling.}  We start \nil{by} comparing, in \Fig{fig:luminosity},
the Johnson V-Band luminosity of the bound satellites to the
backsplash and infalling population of galaxies as well as the
observational data as taken from \citet{Koposov08} and
\citet{Maccio10}, respectively (thin solid line, referred to as
``Maccio' sample'' from now on): these data are a combination
of the volume corrected MW satellite luminosity function
\citep{Koposov08} augmented with information from \citet{Mateo98} and
\citet{Maccio10} kindly provided to us by Andrea Maccio \nil{(personal
  communication)}.  And even though our bound luminosity function
agrees with the Maccio data rather well, we stress that we
included the observational data merely as a reference to guide the
eye. It is not our prime objective to reproduce the MW and/or M31
luminosity function of satellite galaxies with our
simulation. However, a close match (as seen in \Fig{fig:luminosity})
reassures us that our simulation is not too far fetched and that our
method for lighting up subhaloes (cf. \Sec{sec:lightingup}) yields
credible results. The central theme of this paper is the comparison
between the (numerically obtained) infalling and backsplash population
and possibilities to decipher them photometrically.

\begin{table}
\begin{center}
 \begin{tabular}{lclc}
\hline
   \multicolumn{2}{r}{comparison} & & $p$ \\
\hline
\hline
$z=0$\\
\hline
  bound & -- & backsplash    & 0.114\\
  bound & -- & infalling         & 0.190\\
  backsplash & -- & infalling & 0.667\\
\\
$z=1.5$\\
\hline
  bound & -- & backsplash    & 0.051\\
  bound & -- & infalling         & 0.443\\
  backsplash & -- & infalling & 0.417\\
\hline
 \end{tabular}
 \end{center}
 \caption{Kolmogorov-Smirnoff (KS) probabilities $p$ for various
   comparisons of the luminosity functions presented in
   \Fig{fig:luminosity}.}
\label{tab:KS}
\end{table}

To better quantify the differences and similarities between the
respective simulated luminosity functions we applied the
Kolmogorov-Smirnoff (KS) test that provides us with the significance
level $p$ that the null hypothesis that two data sets are drawn from
the same parent distribution; small values of $p\in [0,1]$ show that
the two cumulative distribution functions (i.e. in our case two
luminosity functions) are significantly different.\footnote{We
  utilized the routine \texttt{kstwo()} as described in
  \citet{Press92}.} We \nil{find that the KS probability} that the
backsplash and infalling distributions have been drawn from \nil{the
  same parent} function \nil{is} 67\%. The significance level is \nil{just}
11\% when comparing the backsplash with the bound population and 19\%
when comparing the infalling with the bound satellites. These numbers
and probabilities, respectively, have been summarized in
\Tab{tab:KS}.

In addition to calculating the KS probability $p$ that these
distributions stem from the same parent distribution we also performed
the experiment of randomly drawing $N_{\rm back}$ galaxies from the
infalling and bound sample where $N_{\rm back}$ is the number of
backsplash galaxies. Comparing the resulting down-sampled luminosity
functions again using a KS test we find a probability $p$ of 
$p\approx 0.66$ \nil{when} \nil{comparing the} backsplash \nil{population to
  the} infalling \nil{one} and $p\approx0.12$ \nil{when comparing} the
backsplash \nil{to the} infalling or bound satellites
\nil{population}. All this hints at similarities between backsplash
and infalling satellites whereas the bound population has likely
evolved differently.

\begin{figure}
\noindent
\centerline{\hbox{\psfig{figure=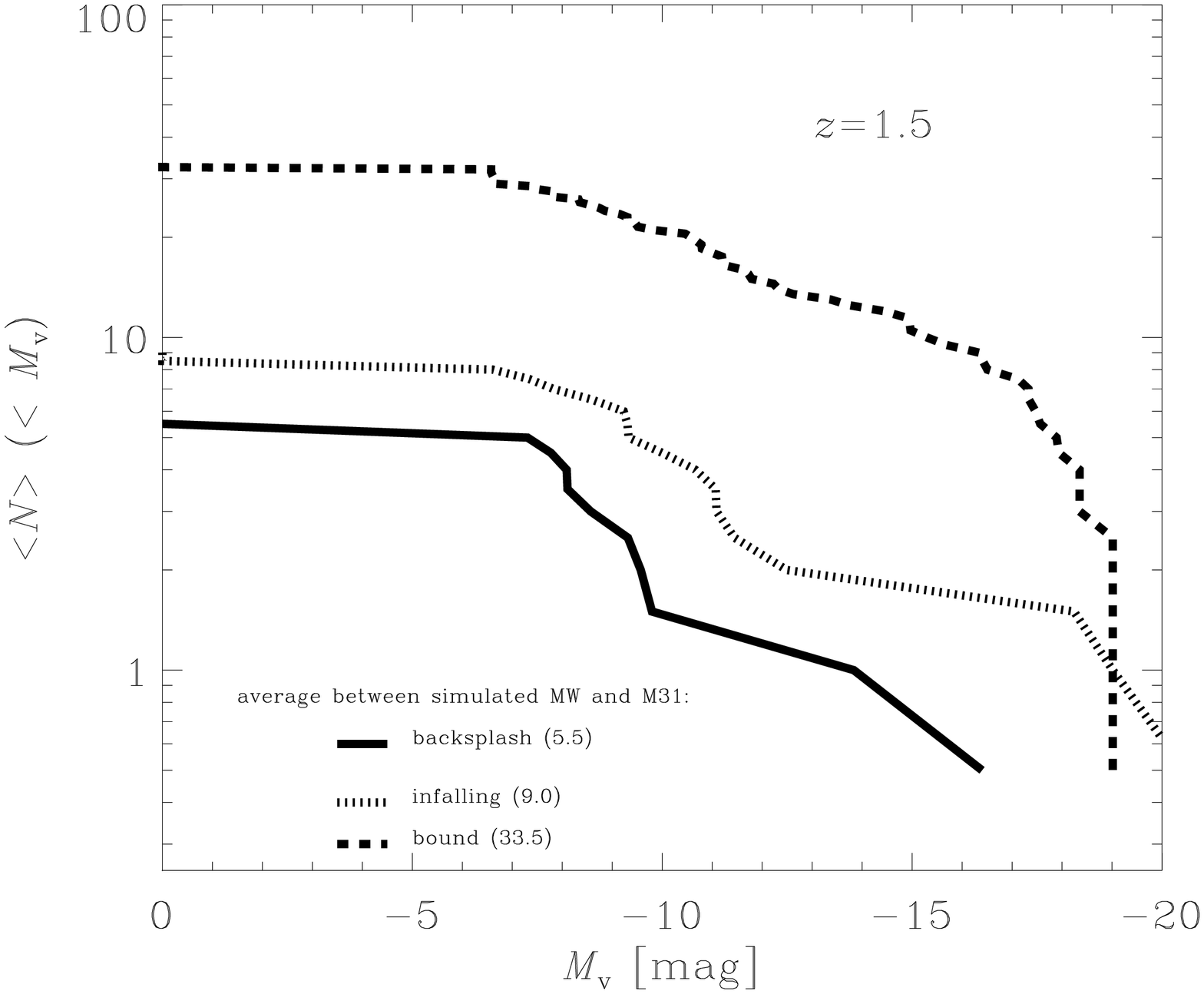,width=\hsize,angle=0}}}
\caption{The luminosity function of subhaloes in the Johnson V-Band at
  redshift $z=1.5$ (i.e. the redshift at which none of the galaxies
  has yet entered their respective host).}
\label{fig:luminosity-z1.5}
\end{figure}

Since all bound and backsplash galaxies themselves \nil{were}, at some
stage, infalling satellites the differences between the bound and
backsplash/infalling luminosity function at redshift $z=0$ should (at
least) be \nil{lessened} when moving to a time where none of the
objects had entered their host, i.e. redshift $z=1.5$
(cf. \Fig{fig:orbits} in \Sec{sec:existence}); the three (cumulative)
distributions of luminosities at redshift $z=1.5$ are presented in
\Fig{fig:luminosity-z1.5}. Performing the same exercises of comparing
the various distributions using the KS statistic (cf. \Tab{tab:KS}
again), we obtain a marginally larger probability for the infalling
population to agree with the bound and backsplash subhaloes. However,
the compatibility between the bound and backsplash is actually
lowered. In that regards we need to stress that the number of
satellites -- despite combining MW and M31 -- is not very large
(especially not for the backsplash population) and hence any
(extensive) statistical analysis has to be taken with care. Therefore,
the probabilities presented here are more indicative of possible
trends rather than providing hard evidence for similarities and/or
differences.

\begin{figure}
\noindent
\centerline{\hbox{\psfig{figure=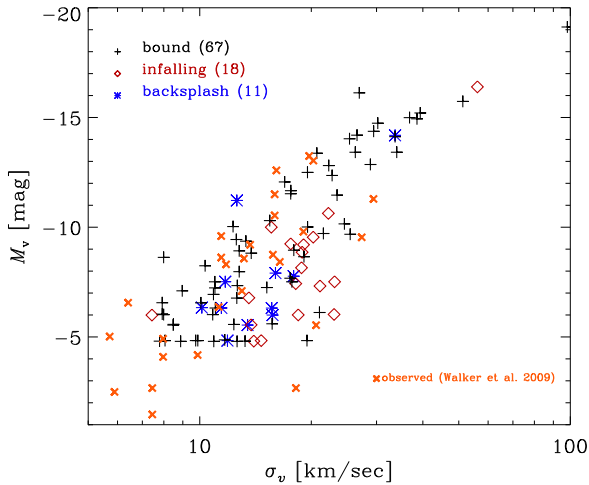,width=\hsize,angle=0}}}
\caption{The relation between Johnson V-Band luminosity and subhalo
  velocity dispersion at redshift $z=0$. The observational data is
  taken form \citet{Walker09}.}
\label{fig:MvSigV}
\end{figure}

Even though our primary motivation is to find a way to
distinguish backsplash from infalling satellites that only utilizes
photometry, we nevertheless present another (observable)
correlation: the luminosity vs. the velocity dispersion $\sigma_v$, in
\Fig{fig:MvSigV}.  As already alluded to above when discussing the
luminosity function, we also added observational data (taken from
\citet[][their Table 1]{Walker09}) simply to guide the eye. While we
also recover the observed trend in our numerical data, the focus
should lie with the infalling and backsplash galaxies. To better
quantify the correlations between $\sigma_v$ and $M_V$ we calculated
the Spearman rank coefficients $R_S$:\footnote{The Spearman rank
  coefficient $R_S$ is a non-parametric measure of correlation: it
  assesses how well an arbitrary monotonic function describes the
  relationship between two variables, without making any other
  assumptions about the particular nature of the relationship between
  the variables \citep{Kendall90}. Its significance $S_S$ is a value
  between 0 and 1 and a small value indicates a significant
  correlation. We use the IDL routine \texttt{R\_CORRELATE()} to
  calculate both these numbers.}  for the observational data it
amounts to $R_S=0.687$ whereas there appears to be a marginally
stronger correlation for our bound satellites of $R_S=0.824$. However,
this ``discrepancy'' is likely due to the different magnitude limits
of both the observational and numerical data, i.e. the two data sets
do only cover the same magnitudes in the range $M_V \in [-13,-5]$.

The respective correlation coefficients $R_S$ for the backsplash and
infalling populations are $R_S=0.373$ and $R_S=0.573$ respectively. While there are
differences in the strengths of the correlation we \nil{find} it
difficult to utilize this interdependence to separate backsplash from
infalling satellites: while the Spearman rank significances $S_S$ are
very close to zero for the bound and observational data (indicating a
reliable determination of the respective $R_S$ value) they are of
order 0.2 for the backsplash and infalling population, probably due to
the small statistical sample.

\begin{figure}
\noindent
\centerline{\hbox{\psfig{figure=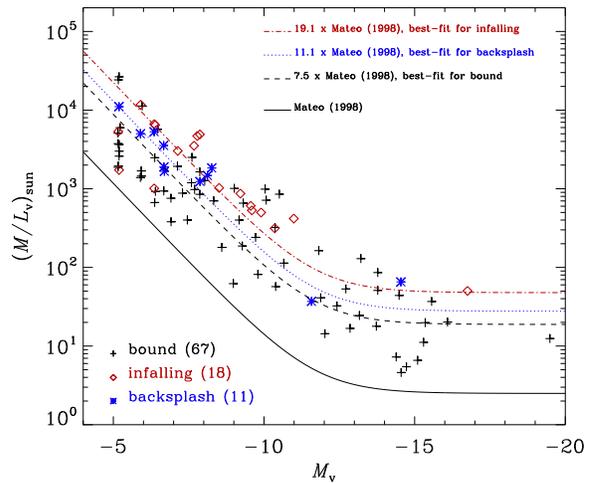,width=\hsize,angle=0}}}
\caption{Mass-to-light ratios (in terms of solar values) as a function
  of V-band luminosity $M_V$. The thin solid line represents the
  observational relation as found by \citet{Mateo98}. The other lines
  are the best-fit curves (with the amplitude as a free parameter) to
  the bound (dashed), backsplash (dotted), and infalling (dot-dashed)
  population, respectively, with the legend listing the respective
  value of the amplitude, too. Note that we used the ``mass $M$ inside
  the visible radius'' as described in the text for this plot.}
\label{fig:ML}
\end{figure}

Above, we showed that while all three populations do follow the same
trend for the $M_V - \sigma_v$ relation (coinciding with the trend
found in observational data), there are nevertheless subtle
differences in the strength of this correlation, especially for the
backsplash and infalling population (cf. the different Spearman rank
coefficients $R_S$). However, the most prominent and well pronounced
difference can be found when studying the mass-to-light ratios $M/L_V$
presented in \Fig{fig:ML} as a function of V-band magnitude $M_V$. We
stress that the mass $M$ used in this plot is actually the
mass within the visible radius of the subhalo; we found the distance
of the farthest stellar particle and used the total mass interior to this
radius as $M$. \citet{Wadepuhl10} already noted that a (substantial)
shift (i.e. $A\approx5.2$) of the observationally determined
analytical relation

\begin{equation} \label{eq:mateo98}
{\frac{M/L}{(M/L)_{\odot}}} = A \left( 2.5+\frac{10^7}{L/L_{\odot}} \right)
\end{equation}

\noindent
is required \citep[][$A=1$ in there]{Mateo98}, which is confirmed by
our data: leaving $A$ as a free parameter and using only the bound,
backsplash, and infalling satellites we find $A=7.5\pm 0.7$ (bound),
$A=11.1\pm 1.2$ (backsplash), and $A=19.1\pm 2.4$
(infalling).\footnote{The fitting to \Eq{eq:mateo98}, i.e. a function
  $M/L (L)$, has been done using IDL's \texttt{CURVEFIT} routine using
  equal weights for the data points $M/L$ vs. $M_V$; the reported
  standard deviations had been returned by \texttt{CURVEFIT}, too.}
These different amplitudes are naturally explained by the differing
histories and (strengths of) interactions with the host. We will see
below in \Sec{sec:baryons} that bound galaxies lost the largest amount
of their dark matter when compared with the other two populations;
infalling satellites are in fact still gaining mass \nil{through
  accretion}. Therefore, taken together with the fact that their
luminosities are nevertheless still similar, we \nil{may infer that}
the mass-to-light ratios \nil{should be significantly} different. This
\nil{notion} opens up the possibility to use the relation presented in
\Fig{fig:ML} to separate the three populations from each
other. \nil{In practice this} requires not only photometric
measurements but also a (proxy for the) mass estimation.

However, using the mass inside the stellar radius also may explain the
differences found in \Fig{fig:ML}: stars in real satellites may be
more compact relative to the dark matter than in our simulation, and
might therefore be less susceptible to tidal stripping (together with
the dark matter inside the ``visible'' radius). This would also
suggest that the differences between our three different populations
might be smaller if the luminous parts of the satellites were more
compact.

We further like to mention that we not only used the mass inside the
stellar radius as a measure for the mass entering the mass-to-light
ratio. We also applied various other definitions, e.g. the total
mass inside the virial radius as well as the mass as determined from
the velocity dispersion under the assumption of virialisation and an
NFW density profile (both at the virial radius and at 15\% of the
virial radius). While the amplitudes $A$ are certainly different when
using different mass estimates, the general trend remains unaltered:
the $M/L$ ratios for the infalling satellites are shifted upwards with
respects to the backsplash population which itself has higher ratios
than the bound subhaloes.

However, we also need to bear in mind one of the subtleties of halo
finding, especially subhalo finding: the definition of mass and the
edge of a subhalo, respectively. While it is straight forward to
define an outer edge for an (isolated) field halo \nil{(usually
  defined as the radius at which the mean interior density drops below
  200 times the critical density)}, the situation is more tricky for
subhaloes: they have to be truncated at the point where their density
profile starts to rise again \nil{due to the} host's \nil{background}
density. Therefore, the same subhalo placed inside and outside of a
host will have different masses due to the nature of (our) halo
finding technique. This explains at least in part the offset in
the mass-to-light ratios for bound/backsplash and infalling galaxies:
the infalling ones have in general higher masses. And part of the gap
between bound and backsplash may be explained by the same phenomenon,
though not all of it; there certainly is no uncertainty that
bound galaxies have lost more mass than backsplash subhaloes.

\begin{figure}
\noindent
\centerline{\hbox{\psfig{figure=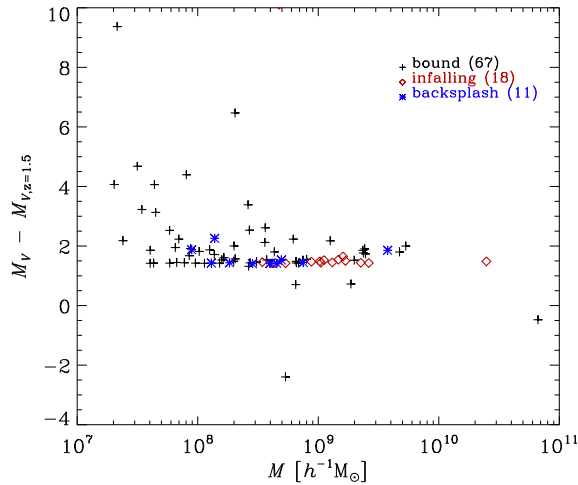,width=\hsize,angle=0}}}
\caption{Difference between Johnson V-Band luminosity at present day
  and redshift $z=1.5$ as a function of present day halo mass $M$.}
\label{fig:MvDiff}
\end{figure}

The differences \nil{between} the luminosities of the populations at
redshift $z=0$ \nil{and} the (marginally) more pronounced similarities
at a time where all populations were still infalling (i.e. redshift
$z=1.5$) \nil{calls} for a closer look at the evolution of
\nil{satellite galaxy} luminosity. To this extent we plot in
\Fig{fig:MvDiff} the difference between the Johnson V-Band luminosity
$M_V$ at redshift $z=0$ and at redshift $z=1.5$ as a function of the
total bound halo
mass $M$ for each galaxy \nil{considered in this paper}, using
different symbols for the different populations (stars for backsplash,
plus-signs for bound, and diamonds for infalling galaxies).

\Fig{fig:MvDiff} now shows several things. For a substantial number of
satellites (especially the backsplash and infalling population) we
only observe a ``constant'' decrease in luminosity of approx. 1.5
magnitudes. However, the luminosity of the bound galaxies drops
significantly -- especially on the low-mass end -- while some of the
higher mass ones gain luminosity. As luminosity is \nil{directly}
linked to stellar content we are left with the question \nil{of} how
these \nil{differences} relate to changes in the stellar population
and/or removal (or gain) \nil{of} star particles \nil{from a
  subhalo}. We study these issues in the following subsection.

Studying luminosities is also closely related to colours, i.e. ratios
of luminosities in different wave-bands. It therefore appears natural
to ask the question - and use the data available to us - to have a closer
look at differences in colours for our three populations. When
plotting the $B-V$ colour as a function of halo mass $M$ (not
presented here) we observe that there are practically no differences
at all amongst the various subhaloes and populations,
respectively. Neither is there are correlation with mass. \nil{Colour
  appears unaffected when categorizing galaxies as bound, infalling or
  backsplash}.

\subsection{The Baryon Content}
\label{sec:baryons}
\begin{figure}
\noindent
\centerline{\hbox{\psfig{figure=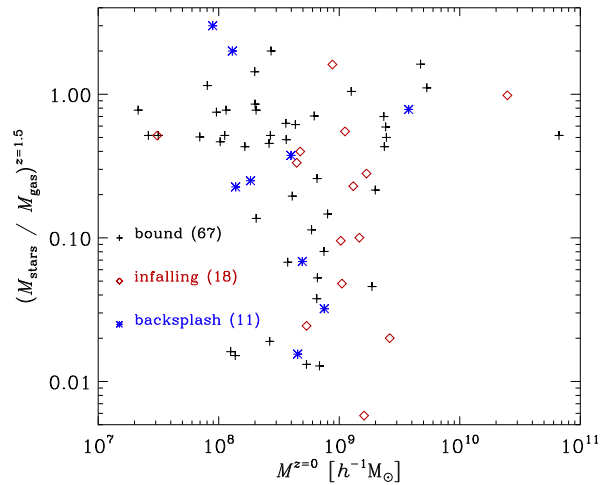,width=\hsize,angle=0}}}
\caption{Ratio of stellar to gas mass at redshift $z=1.5$ as a
  function of present day halo mass $M$.}
\label{fig:fgs-z1.5}
\end{figure}

Before investigating the stellar component \nil{directly} we
\nil{would} like to start with a few words on the \nil{subhalo} gas
content. We \nil{find} that hardly any of the subhaloes under
consideration contain \nil{a significant} gas \nil{content} at
redshift $z=0$. When expressed in terms of the cosmic baryon fraction
the amount of \nil{mass} in gas is of order $<10^{-4}$ for more than
90\% of the subhaloes. However, their stellar mass fractions (again in
terms of the cosmic baryon fraction) is $>10^{-4}$ for all of them
(which is \nil{a direct outcome of  restricting ourselves to a} sample of subhaloes \nil{that} contain
\nil{a stellar component}). The situation though \nil{is} rather
different at redshift $z=1.5$ where we find that all of the
progenitors of the subhaloes not only contained gas but
the fraction of mass in gas \nil{is on} average a factor two higher
than \nil{that} in stars which can be verified in
\Fig{fig:fgs-z1.5}. Note that in this figure we plot the present day
mass on the $x$-axis and the ratio of
stellar-to-gas mass at redshift $z=1.5$ on the $y$-axis. The fact that none of the gas
is left at redshift $z=0$ indicates that either the gas has been
converted into stars \nil{through the process of star formation,} or
the gas has been stripped/removed \nil{through interactions with the
  host halo (e.g. ram pressure stripping)} or other influences prior
to infall. If the former is true we \nil{would expect to} observe an
increase in stellar mass since redshift $z=1.5$, unless there is a
conspiracy at work: the existing stars \nil{may be} stripped at the
same rate as \nil{star formation may} convert \nil{gas} into new
stars, leaving the number of stars unchanged. However, this scenario
is rather unlikely. We \nil{have also} checked for the influence of
the cosmological UV background: \nil{recall} that in our simulations
the thermodynamic properties of the gas are computed in the presence
of a uniform but evolving UV cosmic background generated from
quasi-stellar objects and active galactic nuclei and switched
on at z = 6 \citep{Haardt96}. This \nil{prescription} leads to an
evaporation of gas in objects below a certain mass threshold $M_c(z)$
as given by Eq.~(6) in \citet{Hoeft06}. When plotting the mass
accretion histories of all our subhaloes under investigation here
\nil{and} comparing it to \nil{the} aforementioned formula \nil{in
  \citet{Hoeft06}}, we find that all the backsplash and infalling
galaxies are in fact below the \nil{evaporation} limit. For the bound
objects we find that 1/3 \nil{are}, at redshift $z=1.5$, above that mass
limit. However, \nil{they} also drop below it by $z=0$
(with the odd one remaining above). We therefore conclude that we
should not be surprised to be left with subhaloes that contain hardly
any gas at \nil{$z=0$, due to photo-evaporation by the UV background}.

\begin{figure}
\noindent
\centerline{\hbox{\psfig{figure=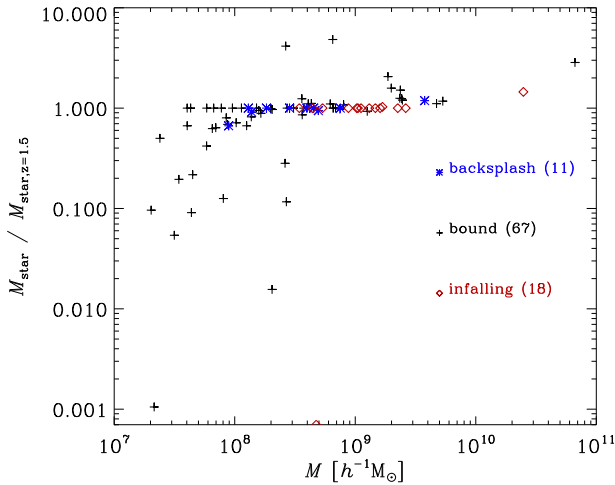,width=\hsize,angle=0}}}
\caption{Ratio of stellar mass at present day and redshift $z=1.5$ as
  a function of present day halo mass $M$.}
\label{fig:MstarRatio}
\end{figure}

The question now is, whether or not we find an evolution of the
stellar component between redshifts $z=1.5$ and $z=0$.  We therefore
plot the ratio of the stellar content at these redshifts as a function
of (present-day) subhalo mass in \Fig{fig:MstarRatio}. We observe that
the backsplash (as well as the infalling) subhaloes hardly lost any
stars since $z=1.5$. Note that our simulations do not model
stellar mass loss and hence the stellar mass remains constant when no
star particles are stripped or newly created. However, this is still
in agreement with the evolution of the luminosity as found in
\Fig{fig:MvDiff}: subhaloes with a constant number of stellar
particles merely evolve passively from $z=1.5$ to $z=0$ due to stellar
ageing. Nevertheless, the lower-mass subhaloes of the bound population
did loose \nil{a} substantial amount of stars while some of the
higher-mass ones gained (or formed) stars. Therefore, \nil{a} picture
\nil{is now} emerging, that while gas has been \nil{efficiently}
stripped, the stellar component remained more or less unaffected -- at
least for the infalling and backsplash population which are of prime
interest in the present study.

\begin{figure}
\noindent
\centerline{\hbox{\psfig{figure=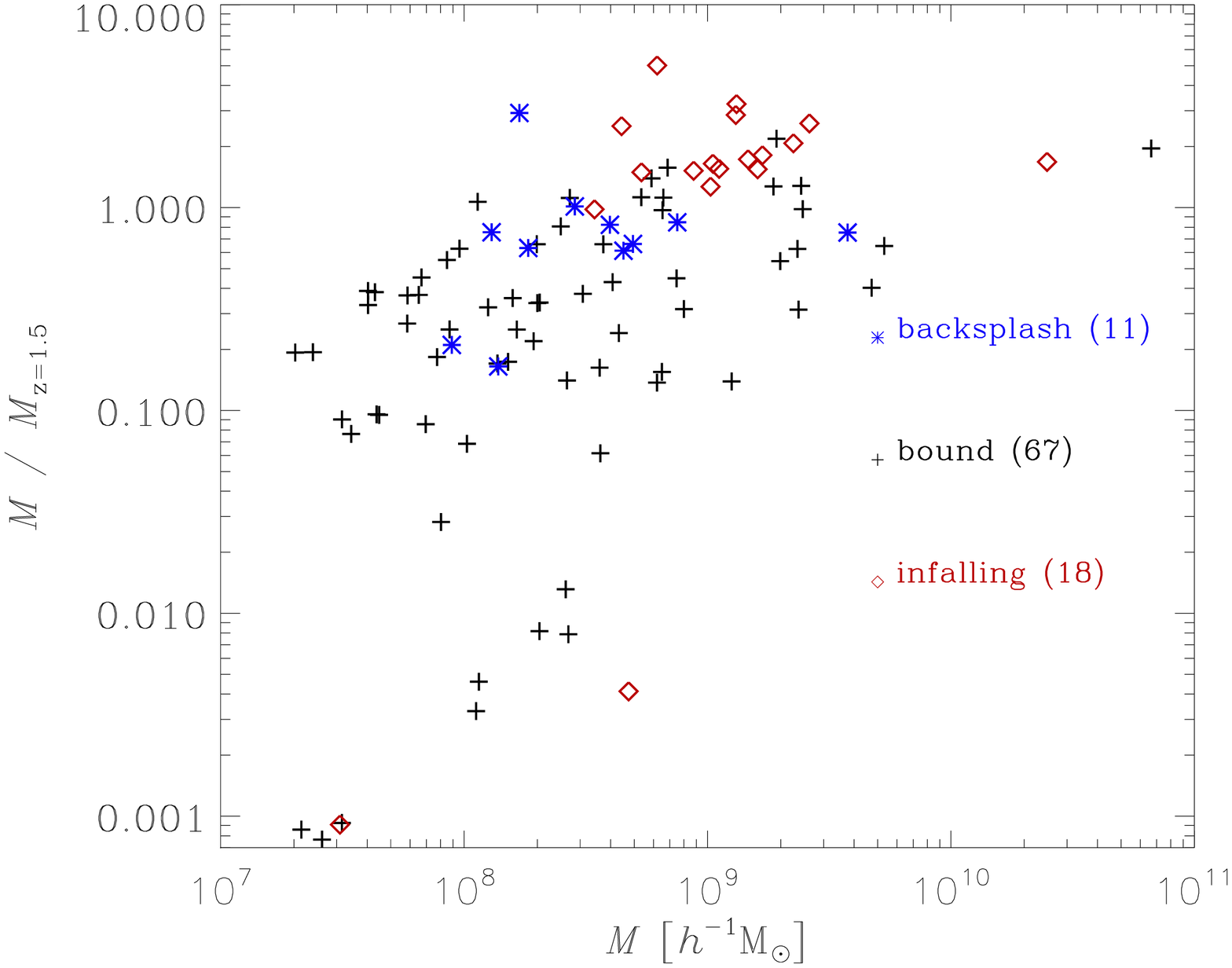,width=\hsize,angle=0}}}
\caption{Ratio of total bound mass at present day and at redshift
  $z=1.5$ as a function of present day halo mass $M$.}
\label{fig:MvirRatio}
\end{figure}

As we expect the stellar component to be concentrated at the centre of
a subhalo, the previous finding on stellar mass loss for bound
subhaloes immediately leads to questions \nil{regarding the nature
  of} mass loss in general. To this extent we show in
\Fig{fig:MvirRatio} the ratio of total bound masses $M$ (again as a
function of today's mass) at redshifts $z=0$ and $z=1.5$. We note that
outside the influence of a host \nil{halo}, subhaloes behave like
field haloes and grow in mass \nil{through accretion processes}; this
is clearly confirmed for the infalling population (\nil{albeit} with
the exception of two objects). We also observe \nil{mass loss via}
tidal stripping, especially for the bound subhaloes. And while
backsplash galaxies may at times loose as much as 40\% of their
original mass \citep{Gill05} we also find the odd backsplash galaxy in
our particular sample that gained mass. Nevertheless, the picture is
more or less clear: even though backsplash galaxies loose mass, their
stellar component remains unaffected. This is not the situation for
bound galaxies that loose both dark matter and stars due to the tidal
interactions with the host, as expected. The picture drawn here
therefore naturally explains the differences in the (amplitude of the)
$M/L_V$ ratios even though for that particular study only the ``mass
inside the visible radius'' has been considered: when using the total
bound mass (not presented here) we recover the same relations amongst
the different subhalo populations with the ratios in amplitude
unchanged; however, the absolute value of the amplitudes is more than
a factor two higher.

\section{Discussion and Conclusions}
\label{sec:concusions}
\nil{In this study} we set out to \nil{examine} the differences of the luminosities of backsplash,
bound, and infalling satellite galaxies in a constrained cosmological
hydrodynamical simulation of the Local Group. \nil{Our} prime question is: \nil{Is it possible} to distinguish these different population by
mere photometry? While we \nil{find} marginal differences in the bound
vs. backsplash/infalling galaxies, the two populations residing in the
outskirts of the host halo appear to have strikingly similar
properties in terms of luminosity. The time backsplash subhaloes spent
under the influence of the host is \nil{therefore} not long enough to affect
the stellar component: they loose mass, but primarily dark matter
and/or gas \nil{particles are stripped - the star particles remain more or less unaffected by the host's tidal field}. Therefore, their luminosity function and luminosities in
general remain akin to the infalling population.

Nevertheless, when allowing for not only photometric information but
also adding ``mass'' to our analysis, we found that the mass-to-light
ratios (as a function of magnitude) are \nil{significantly} higher in
infalling than in backsplash galaxies, which are in turn both higher
than for bound satellites. Fitting the observationally determined relation
presented in \citet{Mateo98} for $M/L_V {\rm \ vs.\ } M_V$ by leaving
the amplitude as a free parameter we find differences in the amplitude
of a factor of 1.5 \nil{and} 2.5 between backsplash and bound and
infalling and backsplash galaxies, \nil{respectively}. \nil{We note,}
however, \nil{that} part of this shift can be explained by (our method
of) halo finding \nil{and certain endemic} limitations when comparing
field and subhaloes. \nil{The radial extent of a} subhalo has to be
truncated due to the embedding within the host's background and hence
has a lower mass than in the case when \nil{the same subhalo is found}
in isolation \nil{(i.e. exterior to a host halo)} even though we
explicitly only considered the ``mass inside the visible radius'' for
this particular part of the investigation. We also need to acknowledge
that the original relation had to be shifted by a factor of 7.2 to
bring it into agreement with our numerical data\footnote{Note that
  \citet{Wadepuhl10} also required a shift by a factor of 5.2 for
  their simulation data.} which nevertheless is not the prime target
of the present paper and its explanation left to a future study,
respectively.

Even though there still remains a lot to be quantified, we believe
that this difference may provide a new window on distinguishing
between infalling and backsplash galaxies that could be applied to
observational data. Its origin is readily explained by the fact that
while backsplash and bound galaxies both lose mass the mass loss is
greater for bound than for backsplash galaxies; therefore, if the
stellar population is unaffected (as found in our simulations) we
observe an enhanced decrease in the mass-to-light ratios for bound
galaxies and -- more importantly for our purposes -- a decrease when
comparing infalling against backsplash.

However, the apparent discrepancy between the simulation presented and
used here and the observational data yet remains unexplained. It could
be possible that stars in real satellites are more compact relative to
the dark matter than in our simulation, and might therefore be less
susceptible to tidal stripping (together with the dark matter inside
the ``visible'' radius). But this would also suggest that the
differences between our three different populations might be smaller
if the luminous parts of the satellites were more compact closing the
aforementioned ``window'' again. In that regards, we remind the reader
that we not only used the mass inside the stellar radius as a
measure for the mass entering the mass-to-light ratio. We also applied
various other possibilities, e.g. the total mass inside the virial
radius as well as the mass as determined from the velocity dispersion
under the assumption of virialisation and an NFW density profile (both
at the virial radius and at 15\% of the virial radius). While the
ratios of the $M/L$ curves are certainly different when using
different mass estimates, the forcited trend remained unaltered.

A closer inspection of \Fig{fig:orbits} reveals that most of the
backsplash galaxies fell into their host at approximately the
same time. When studying the distribution of infall times (not shown
here though) there appears to be a continuous infall of bound galaxies
whereas the backsplash objects all cluster at about redshift
$z\approx0.55$. As pointed out by several other authors recently,
subhaloes may have the tendency to fall into (Milky Way like) hosts in
groups \citep[cf.][]{Klimentowski10, Li09, DOnghia08,
  Li08subgroups}. Hence could it be that all our backsplash galaxies
are part of a larger group? We explicitly checked for \nil{this
  conjecture} by studying their 3D orbits and cannot confirm it: our
backsplash galaxies come from various directions yet fall in at a
similar redshift. However, we also need to acknowledge that these
directions are not random but rather correlated -- however, this has
been studied in detail in a companion paper \citet{Libeskind10infall}.

\section*{Acknowledgements}
AK is supported by the Ministerio de Ciencia e Innovacion (MICINN) in
Spain through the Ramon y Cajal programme and further acknowledges
support by the Ministerio de Education (MEC) grant AYA
2009-13875-C03-02. SRK acknowledges support by the MICINN too under
the Consolider-Ingenio, SyeC project CSD- 2007 -00050. We thank DEISA
for granting us supercomputing time on MareNostrum at BSC and in SGI-
Altix 4700 at LRZ, to run these simulations under the DECI- SIMU-LU
and SIMUGAL-LU projects. We acknowledge support of MICINN through the
Consolider-Ingenio 2010 Programme under grant MULTIDARK
CSD2009-00064. We also thank ASTROSIM for giving us different travel
grants to visit our respective institutions.  GY acknowledges
financial support from MEC (Spain) under project AYA 2009-13875-C03-02
and the ASTROMADRID project financed by Comunidad de Madrid. We thank
Andrea Maccio for kindly providing us with the data of the observed
luminosity function (average of MW and M31).

\bibliography{archive} \bsp

\label{lastpage}

\end{document}